\documentstyle[12pt,aasms4]{article}

\slugcomment{Submitted to the Astronomical Journal}
\begin{document}

\title{Far-Ultraviolet Spectroscopy of Star-Forming Regions in Nearby Galaxies:
Stellar Populations and Abundance Indicators
\footnote{Based on observations made with the NASA-CNES-CSA Far Ultraviolet 
Spectroscopic Explorer. FUSE is operated for NASA by the Johns Hopkins 
University under NASA contract NAS5-32985.}}

\author{William C. Keel}
\affil{Department of Physics and Astronomy, University of Alabama, Box 870324,
Tuscaloosa, AL 35487; keel@bildad.astr.ua.edu}

\author{Jay B. Holberg}
\affil{Lunar and Planetary Laboratory, University of Arizona, Tucson, AZ 85721; 
holberg@argus.lpl.arizona.edu}

\author{Patrick M. Treuthardt}
\affil{Department of Physics and Astronomy, University of Alabama, Box 870324,
Tuscaloosa, AL 35487}

\begin{abstract}
We present FUSE spectroscopy and supporting data for star-forming regions
in nearby galaxies, to examine their massive-star content and explore the
use of abundance and population indicators in this spectral range for
high-redshift galaxies. New far-ultraviolet spectra are shown for 
four bright H II regions
in M33 (NGC 588, 592, 595, and 604), the H II region NGC 5461 in M101,
and the starburst nucleus of NGC 7714, supplemented by the very-low-metallicity
galaxy I Zw 18.
In each case, we see strong Milky Way absorption
systems from H$_2$, but intrinsic absorption within each galaxy is
weak or undetectable, perhaps because of the ``UV bias" in which 
reddened stars which lie behind molecular-rich areas are also heavily reddened.
We see striking changes in the stellar-wind lines from these populations
with metallicity, suggesting that C II, C III, C IV, N II, N III, and
P V lines are potential tracers of stellar metallicity in star-forming
galaxies. Three of these relations - involving N IV, C III, and P V -
are nearly linear over the range from O/H=0.05--0.8 solar.
The major difference in continuum shapes among these systems is
that the giant H II complex NGC 604 has a stronger continuum shortward
of 950 \AA\  than any other object in this sample. Small-number statistics
would likely go in the other direction; we favor this as the result of
a discrete star-forming event $\approx 3$ Myr ago, as suggested by
previous studies of its stellar population.  
\end{abstract}


\section{Introduction}

The history of galaxies is, in large part, the history of star formation.
Massive stars play key roles both as highly visible tracers of star
formation and as players in altering surrounding star formation
and both energy and chemistry of the interstellar medium. It is
these stars which dominate the observed properties of actively
star-forming galaxies. The massive part of stellar populations
is most clearly observed in the ultraviolet, where their energy
distributions peak and competing light from cooler stars is minimal,
as long as the foreground extinction allows escape of enough of this
radiation. In such environments,
ultraviolet studies of star-forming regions have proven fruitful in
understanding these populations.
The recent opening of the far-ultraviolet window, between Lyman $\alpha$
and the Lyman limit, for deep observations, allows study of massive hot
stars in a range where they fully dominate the spectrum. This relatively
narrow band contains an embarrassment of spectral riches, with numerous
lines from stellar winds as well as interstellar material both atomic
and molecular. These include the strong and highly-ionized lines of
O VI and the unique ability to measure cold H$_2$. In addition, this
piece of the spectrum is accessible for high-redshift galaxies, at least
for composite samples where the Lyman $\alpha$ forest can be averaged
adequately, allowing the possibility of direct comparisons of stellar
populations over a large span of cosmic time.

The very sensitivity of the far-UV light to star formation and reddening
makes it a purer probe of some properties of star-forming regions than
observations at longer wavelengths.
Since only short-lived stars contribute significantly in the far-UV
range, the details of star-forming history should matter only for
very brief bursts (such as might be found in individual H II regions,
but are less likely on galaxy scales). This makes the far-UV spectrum
more sensitive to the stellar population itself than to its history.
Furthermore, although the extinction is high, its differential effect
across the far-UV band is modest, and paradoxically the effective
reddening to the stars we see is smaller than found at longer wavelengths.
In observing stars intermingled with highly structured dust distributions,
the ``picket-fence" effect (Heisler \& Ostriker 1988) means that most 
of the stars are so
reddened as to make no significant contribution in the deep ultraviolet;
all the stars we see are only lightly reddened. This also reduces the
effects of the forest of H$_2$ absorption features because of the mixing
of molecular gas and dust.

Observational data on nearby galaxies, before 
the {\it Far-Ultraviolet Spectroscopic Explorer} (FUSE), 
were limited to a handful
of star-forming systems. Four starburst galaxies were observed using 
HUT on {\it Astro-2} (Leitherer et al. 1995, 2002), largely to measure escaping 
radiation in the
Lyman continuum, which provided initial data for comparison with
models based on stellar spectra dating back to {\it Copernicus}.
The strongest features fall into two blends near 970 (Ly $\gamma$ + C III)
and 1030 \AA\  (O VI + Ly $\beta$ + C II).
A {\it Voyager 2} observation of M33, with some spatial resolution
in one direction, was analyzed by Keel (1998), showing that its
far-UV continuum is virtually identical to those of the powerful
starbursts, and that NGC 604 is bluer in this range than the overall
disk. These data also suggested significant Lyman $\alpha$ emission,
with spatial profile suggesting an origin in the diffuse ISM rather than
giant H II regions. Ironically, until the availability of FUSE
observations, the richest information on galaxies shortward of Lyman
$\alpha$ came from objects at high redshift, particularly composite
spectra of Lyman-break galaxies (Steidel et al. 2001, Shapley et al. 2003).
To enable comparisons between local, well-studied star-forming systems
and these powerful, young objects, we have undertaken a series of FUSE
observations of star-forming regions innearby galaxies. We present here
the analysis of these spectra in the context of their stellar populations
and systematic changes with metallicity. A companion paper (Keel, Shapley,
\& Steidel 2004) considers the comparison with composite spectra of
Lyman-break galaxies.

\section{Observations}

\subsection{Object selection}

The FUSE targets were selected for known high UV flux and hence
low reddening, and to span
a range of metallicity. The regions targeted in M33 include the brightest
H II regions in the mid-UV, and span a radial range within the disk
from 1.6-4.5 kpc (see Fig. 1). They also include examples with
a single dominant star cluster and with multiple clusters or
more diffuse associations (also illustrated in Fig. 1 using HST
data at 1700 \AA\ ). None of the objects in M33 or M101 achieves
``super star cluster" (SSC) status, if we follow common usage in requiring 
not only high luminosity and stellar mass (Melnick, Moles \& Terlevich 1985)
but that the stars be concentrated into a single clump on
10-pc scales as noted by such studies as that of
Meurer et al. (1995). While NGC 604 has the right luminosity,
its stars are widely spread throughout a 100-pc region including
multiple clumps
(Hunter et al. 1996). In the more
distant systems, the FUSE aperture samples multiple regions; Fig. 2
shows the bluest archival HST imagery with the aperture superimposed.
In NGC 7714, there are $~\approx 10$ luminous SSCs, comparably luminous
in the near-UV,  within the aperture,
more clearly shown in the inset to Fig. 2. In each of these
cases, essentially the entire star-forming region fits within the
aperture, an important desideratum in comparing these observations
with global measurements of distant galaxies.

For each of these objects, there are abundance measurements from the
traditional optical emission lines, which we use as an widely-applicable
tracer of the abundances of recently-formed stars as well.
The M33 objects span much of its disk's abundance gradient,
while NGC 5461 and the nucleus of NGC 7714 were selected to
sample lower and higher metallicity. We also analyze the
summed archival FUSE spectrum of I Zw 18 from Aloisi et al. (2003),
as a comparison with the lowest known gas-phase abundances.

Table 1 includes our adopted values from the literature for
[O/H], the best-measured of the abundances due to the optical lines
from multiple ionization stages of oxygen and its importance as
a coolant.

\subsection{FUSE Spectroscopy}

The FUSE optical system and detectors are described by Moos et al. (2000).
Four primary mirrors are used to feed independent detectors optimized
for subsections of the far-UV band; maintaining the coalignment of these
optical systems is an important operational issue. Each detector has
a distinct wavelength calibration, but for our purposes, we are
not pushing the resolution limit of FUSE and can combine the various
data segments in 0.1-\AA\  pixels.

As shown in Figs. 1 and 2, the FUSE pointing positions were set at the
midpoint of the stellar distributions from available UV images.
The large 30" aperture was specified for the M33 and M101 H II
regions at the outset, both for flux integrity in alignment of the four
instrumental channels and 
to include essentially all their starlight,
and adopted for NGC 7714 as well when the performance penalty
for using a smaller aperture became clear during the mission. 
The requested center coordinates and total exposure times are listed
in Table 1.

Of these objects, NGC 604 is the brightest in the far-UV range by
a factor 10, and will thus play a continued role in our knowledge of
stellar populations in this spectral range. In each of these
spectra, the useful resolution is limited by the signal-to-noise
ratio, dictating the wavelength binning to detect features of
interest. For NGC 604, we can work at 0.1-\AA\  binning, for a velocity
spacing typically 30 km s$^{-1}$. Only in this object does it
(marginally) matter that it is not a point source, with starlight 
coming from most of the aperture area. This smearing along the dispersion
axis contributes a broadening of about 60 km s$^{-1}$.

The spectra were processed through the FUSE pipeline; we inspected
the two-dimensional spectra to verify data quality and background
subtraction, and to note such
effects as the ``worm", a shadowing of the detector by a repeller wire, 
near the red end of some spectra. Each of
our targets was observed during two orbits, so we could compare
each orbit's summed spectra as a check on the errors at each wavelength.
Loss of flux due to the ``worm" does occur in some of the spectra
around from 1170--1180 \AA , a range which does not affect any of the
features we analyze. Similarly, there are spurious emission features
near 1044 and 1169 \AA\  which are second-order scattered solar
features, which do not seriously confuse features of interest.
For clarity of illustration, we interpolate across terrestrial airglow
features in Ly $\beta$, Ly $\gamma$ and O I at the highest resolution
before further averaging and display.

We also use the summed archival spectrum of I Zw 18 described by
Aloisi et al. (2003), who kindly provided it in electronic form.
We compare the spectra of NGC 604 and I Zw 18 at 0.1-\AA\  resolution
in Fig. 3, showing foreground absorption from H$_2$, atomic absorption
lines in the foreground and in these galaxies, and prominent
stellar-wind lines.

\subsection{Supporting Data}

We draw on additional data from a variety of sources for these
well-observed objects. IUE spectra in the large $10 \times 20$-arcsecond
apertures give mid-UV spectra in roughly matched apertures; we
have collected spectra from the archive, rejecting those with
obviously discrepant pointings or flux levels from the combined
spectra.

HST imaging is invaluable in examining the stellar populations, resolving
the brightest stars in each of these associations. Archival WFPC2 data exist
for each of the M33 regions in F170W plus longer-wavelength filters,
so we can assess the stellar statistics at least into the mid-ultraviolet.
The best UV images of NGC 604 were taken in support of a slitless-spectroscopy
program, using the STIS NUV-MAMA detector working at 1800-2700 \AA\ .

Each of these stellar collections ionizes a substantial surrounding region.
H$\alpha$ data (as in Bosch et al. 2002) show structure extending 
200-800 pc. All four of the M33 regions fit their description of
an evolved H II region, one in which star formation has been long
enough to generate extensive filaments and loops driven by stellar
energy input; they estimate ages $> 4$ Myr for all except NGC 604, 
at 3 Myr, for the current episode of star formation 
(matching the conclusions of Gonzalez Delgado \& Perez 2000).

Stars which contribute strongly in the far-UV have spectral types
B0 and earlier, for ages up to 15 Myr. However, the continuum
slopes of the shorter-lived O stars vary only slightly across
the far-UV range, so the continuum offers little signature of the
recent star-forming history (Robert et al. 2001). To this point,
the far-UV line behavior is not well enough calibrated to infer
the star-forming history independently.

\section{Interstellar Absorption Features}

The far-UV range is rich in narrow absorption lines from the ISM,
both ionic and molecular. The spectrum of NGC 604 is especially complex,
with over 100 detected and identified transitions. Foreground H$_2$ in the
Milky Way is so prominent in all these objects that it must be carefully
accounted for in identifying intrinsic absorption features.
For the strongest Fe II lines, we detect not only the M33 ISM (at about
-200 km s$^{-1}$), but the
foreground high-velocity cloud structure near -370 km s$^{-1}$ described
by Wakker et al. (2003).

\subsection{Atomic features}

Numerous atomic absorption features are seen in these spectra,
with stronger absorption in most cases from foreground
Milky Way gas.
Their strengths in the integrated spectra will reflect the distribution
of gas toward the stellar associations weighted by contribution to the
far-UV flux; for these extensive distributions of stars,
derived column densities should be regarded as characteristic values.

\subsection{Molecular hydrogen}

The Lyman and Werner band systems of H$_2$ are at once an advantage and
a nuisance in far-UV spectroscopy, detected in almost every line of
sight crossing a significant distance in our own galaxy's ISM. A total
of 56 such lines are individually detected in NGC 604.
The H$_2$ features in these M33 data have also been analyzed by Bluhm et al. (2003), 
who found that H$_2$ absorption at the velocities appropriate for disk
gas in M33 is weakly detected in NGC 588, 592, and 595, at column
densities N(H$_2$)= $ 10^{16}- 3 \times 10^{17}$ cm$^{-2}$, but only an upper
limit $\approx 10^{15}$ could be derived
for NGC 604, despite the higher data quality and comparable
column densities from atomic lines, with all the values
approximate due to the lines being too weak for independent measurement of the
Doppler-width $b$ parameter. Bluhm et al. also present 
instructive simulations
on the difficulty of deriving unique column densities when the
background source consists of stars spanning a range of column density and
position within the spectroscopic aperture. For example, the nondetection
in NGC 604 might result from patchy extinction,
correlated with the H$_2$ distribution, and the far-UV background flux being
dominated by those stars with the shallowest H$_2$ absorption. This will
act in addition to any physical effects related to the radiation field
and shocks in the vicinity of such an active star-forming region.
These regions are certainly associated with large concentrations of
molecular gas; all three of them within the survey area of Engargiola et al. 
(2003) are associated with molecular clouds (NGC 592, 595, and 604 with
their clouds 40, 32, and 8, respectively). Their H$_2$ masses are
estimated in the range $1.2-4 \times 10^5$ solar masses.

In spite of these issues in physical interpretation of molecular absorption,
the empirical results are directly comparable to what we may see in
high-redshift systems, as this spectral region becomes accessible for
at least composite study in Lyman-break galaxies (Shapley et al. 2003).
We can use both the sets of H$_2$ line strengths seen from Galactic gas
in these spectra, and synthetic absorption spectra, to ask what 
signatures of molecular absorption remain at modest spectral resolution.
The least confusion with other spectral features for modest H$_2$
column densities occurs for blends of features near 1005 and 1072
\AA . The wavelengths and shapes of these blends depend on the
spin-level populations in the gas, generally shifting to longer
wavelengths for warmer gas (McCandliss 2003). 

We have re-examined the NGC 604 spectrum for evidence of associated H$_2$
absorption. The ability to see such features at low redshift is limited
largely by blending with foreground Galactic features, since the spacing
between multiplet members provides ample opportunity for overlap. For
atomic absorption species, we find a difference between foreground
and M33 absorption typically 220 km s$^{-1}$, which gives wavelength
shifts very close to the spacing of low-order members of the H$_2$
multiplets near 963, 982, 1002, and 1014 \AA\ . As a result, only
the bluest members of these multiplets could be securely detected from
gas near the velocity of NGC 604 in the presence of the much stronger
foreground gas. To guard against unrecognized
atomic features, we also require that a putative NGC 604 feature
not be seen in the I Zw 18 spectrum at the corresponding {\it emitted}
wavelength. Compared to 56 H$_2$ lines seen from foreground Milky Way
gas, only two potential detections resulted for NGC 604. These are the
transitions at rest wavelengths 1001.69  ($5,4 \rightarrow 13,1$)
and 1008.38 ($2,3 \rightarrow 8,0$) at equivalent widths of roughly
0.02 and 0.08 \AA\ , respectively. The 1008 line, with better S/N
ratio, appears substantially broader, with a profile compatible with
that of the other narrow features. If accurate, these detections
are insufficient to derive the absorbing column density, lacking information 
on the intrinsic Doppler broadening and level populations. If 
we adopt the ``optimum" $b$-value 10 km s$^{-1}$ for the 
NGC 588 and 592
detections from Bluhm et al., and incorporating the H$_2$OOLS template
models described by McCandliss 2003, we derive 
N(H$_2$)=$3 \times 10^{17}$ cm$^{-2}$ from this single transition and assuming
the same mix of level populations as in the foreground gas. Hotter
molecular gas could give comparable absorption at column densities
several times lower, close to the Bluhm et al. limit.

Lines arising in the states $J = 0-3$ are seen in the foreground 
Milky Way gas, 
well below the saturated level that 
drove Rachford et al. (2002) to use
detailed profile fitting. As in their work,we also find significant changes in
derived N(H$_2$) between various absorption bands.
We follow them in obtaining a characteristic kinetic temperature
from the level populations in $J=0,1$ as
$$ T_{01} = {{74 {~ \rm K}} \over{\log N(0) - \log N(1) +0.954}} .\eqno{(1)}$$
For the NGC 604 foreground H$_2$ spectrum, we obtained $T_{01} = 113$ K, 
at the high end of values seen by Rachford et al. for dense and self-shielded
environments but reasonable for the more diffuse ``intercloud" ISM. 
The populations in $J=0,1$ are taken to represent a thermal (collisional)
temperature, with ``excess" absorption from higher $J$ representing
radiative excitation, so that only the lowest levels are useful in
deriving the thermal environment of the molcular gas.

Similar issues affect the strongest H$_2$ lines in I Zw 18,
albeit with less overlap from Milky Way features because of the
larger velocity shift. From 
the stacked spectrum, Aloisi et al. (2003) set an upper limit
of N(H$_2$) = $5 \times 10^{-14}$ cm$^{-2}$ against the emerging
continuum.


\section{Stellar Content}

\subsection{Stellar winds}

Most of the stellar signature in far-UV spectra occurs in the wind lines,
which dominate this spectral range thanks to the numerous resonance
transitions of metals. Both the wind lines and the evolutionary tracks
when winds are important should have strong metallicity
dependences, since the radiation pressure driving the
winds acts largely through the opacity of heavy elements. This behavior has been
shown, albeit sometimes in complex ways, upon comparison of stellar
spectra from the Milky Way and Magellanic Clouds. Leitherer et al. (2001)
examined wind lines in the mid-UV, notably C IV, from SMC to Galactic
abundances, confirming a significant metallicity dependence in its
strength but noting than ionization states other than the dominant
ones can behave differently as secondary abundance effects shift the
ionization balance in the winds (as seen in Si IV, which is stronger 
in LMC stars than the Milky Way). These changes are most pronounced
for the brighter luminosity classes.  
Among the most luminous stars, those in the Magellanic Clouds have
smaller wind velocities even for comparable depth (e.g. Kudritzki \& Puls 2000).

While photospheric lines in the far-UV range are weak enough to scale
directly with abundances, they are either extremely weak or overlain
by wind lines. Lamers et al. (1999) suggest that line blanketing is
a more secure route to the stellar abundances. From the
comparison by Robert et al. (2001), Si IV $\lambda 1122/8$ are
high-excitation photospheric lines that should serve as useful indicators
of the stellar population, independent of winds.

In interpreting the O VI profiles from galaxy-scale systems such
as NGC 7714, Gonzalez Delgado et al. (1998) caution from HUT
spectra that interstellar absorption from large-scale outflows
can significantly overlap wind lines, specifically between
Ly $\beta$ and O VI $\lambda 1032$.

Even FUSE-based spectral syntheses in the far-ultraviolet extend only
to 1000 \AA\  (Robert et al. 2003), so we continue our empirical approach in 
comparing
systems of various properties to seek differences linked to composition 
or history. We can, however, be guided by available syntheses of some of the
prominent blends of features, such as the Ly $\beta$-O VI - C II
regions from 1025-1038 \AA\  (Gonzalez Delgado et al. 1997). They find
that Ly $\beta$ and C II absorption arise in B stars, at the cool end of
far-UV contributors, while O VI P Cygni profiles come from the most
massive stars. High spectral resolution is crucial to separating the
interstellar absorption contributions in each case; for Ly $\beta$,
hydrogen column densities $N_H > 10^{21}$ cm$^{-2}$ lead to blending of
stellar and interstellar components. Robert et al. (2001) show that
O VI is very weak at SMC metallicities, and that the youngest population
can be diagnosed from the presence of C II $\lambda 1176$ 
and C IV/N IV $\lambda 1169$ which are specific to hot O stars.
At Magellanic Cloud abundances, the S IV $\lambda 1063/73/74$ lines have
a wind contribution only from supergiants, and P V $\lambda 1118/25$
exhibit similar behavior.

To show the some of the spectral differences seen with metallicity,
Fig. 4 compares the spectra of NGC 604 and I Zw 18, now with the
strongest Galactic H$_2$ features removed by fitting Gaussians
or Voigt profiles (for the stronger lines), and plotted in the
emitted wavelength frame. Fig. 5 is a similar comparison of
NGC 7714 with NGC 604, illustrating the yet stronger features
seen at near-solar metallicity.

The wind lines seen in these systems may be described as follows.
For the fainter objects with short exposures (NGC 588, 592, 595, 5461,
7714) the data have been smoothed by typically 0.7 \AA\  in making
these assessments.

NGC 604 (Fig. 4): Combining both members of the O VI doublet to reduce
confusion, there is a broad wind absorption reaching 2800 km/s
and probably blending with Ly $\beta$. O VI $\lambda 1037$,
C III $\lambda 977$, and N II $\lambda$ 1083 have nearly
black cores.
A distinct detached absorption may be present in C III from
1600-2000 km/s. There may be a broad wind feature in N III to $\approx 3000$
km/s. P V is weak except for possibly photospheric components
near zero velocity. The emission sections of P Cygni profiles
are strong in O VI $\lambda 1037$, N II $\lambda 1082$, and
N III $\lambda 991$. 

NGC 588/592/595: These spectra have wind features as much like one
another as their S/N ratio can tell, both individually and
when averaged. Compared to NGC 604, the cores of C III, O VI, and N II are
shallower, with residual intensities 10-40\%. As noted below, the
continuum level in NGC 604 is higher than in any of the other
systems below $\approx 950$ \AA .

NGC 5461: a trough occurs in N IV to about 1000 km s$^{-1}$. There is a single
well-defined trough in C IV to 600 km s$^{-1}$. No feature is obvious
in N III or S III. Broad absorption is seen in both O VI lines
to about 1000 km s$^{-1}$.

I Zw 18: There is at most weak O VI absorption between 1700-2800 km s$^{-1}$,
seen only in the $\lambda 1037$ line. P Cygni emission is absent
in O VI, N II, and N III by comparison with NGC 604.
Essentially no blueshifted troughs occur for N III. Two features may be
present for C III from 0-1200 and 1700-2000 km/s, but blending
with saturated interstellar C II is an issue. P V is very weak.

NGC 7714: This nucleus has the strongest wind features in our study.
Both O VI lines are strong, extending to 900 km $^{-1}$. 
The P V lines at $\lambda \lambda 1118,1125$ are broad
and blueshifted suggesting a wind contribution, which is plausible
for non-supergiants at its metallicity. N III 991 may have a
broad wind component blending with the interstellar O I lines at about 1000 
km s$^{-1}$. The core of
C III is blended with O I but shows a wind trough to beyond
1000 km s$^{-1}$. 

For high-redshift objects, it will be difficult or impossible to
separate wind and interstellar contributions to some of these lines.
For purposes of comparison, we have generated simple equivalent-width
values with respect to the adjacent (pseudo)continuum for
lines in clean parts of the spectrum, typically spanning a
5-\AA\  line region, and tabulate
these in Table 2. Airglow emission contaminated the
fainter M33 regions (combined in the table as ``M33 avg")
too strongly for a reliable measurement of
N III $\lambda 991$.
For the N II and N III lines we list both the absorption and net
equivalent widths (in parentheses), for use when the components are not 
resolved. We include values for C IV $\lambda 1548,1550$ from
summed IUE low-dispersion spectra, as these values were taken
with a large enough aperture to sample most of the stellar population
in a way much like the FUSE aperture.
Errors on the FUSE value are typically 0.2 \AA\  for such
broad features, as reflected in the limits for some lines in
I Zw 18, while the C IV errors are closer to 1 \AA . 
Each of these lines shows a strong trend with
emission-line metallicity (Fig. 6), showing that they
do in fact have potential use as abundance indicators. The
wind lines measure stellar values directly, in contrast to the
large regions of the interstellar medium sampled by emission-line
techniques. These relations are quite sensitive in the sub-solar
regime of particular interest for the evolution of galaxies
at $z> 3$. The equivalent widths measured for N IV, C III, and P V 
vary almost linearly with abundances. Our results indicate that
these strong features can be used as metallicity indices for
high-redshift stellar populations, in regimes for which the
optical emission lines lie in infrared bands of high atmospheric
emission and absorption, and can be a valuable tool for approaching
the chemical evolution of galaxies in the range $z=3-4$.

For ease of use, we present least-square quadratic fits to the
data in Fig. 6, along with the derived constants needed to
invert these fits with line equivalent widths as the independent
variables. These interpolation curves are shown in Fig. 6, and the
numerical values are found in Table 3. The O/H ratio is in solar
units as found from emission-line analysis (as cited in Table 1, updated to 
the ``new" solar abundance scale with 12 + log (O/H)=8.69), and all equivalent
widths are in \AA\  in the emitted frame. For each line, the data have
been approximately fitted in the form $ {\rm EW} = a_1 + a_2 {\rm (O/H)} 
+ a_3 {\rm (O/H)}^2$.
Similarly, an inverse fitting function for metallicity derived from each
line is obtained from the quadratic formula, with constants
tabulated as ${\rm (O/H)} = c_1 + c_2 (c_3 + (c_4 {\rm EW}))^{1/2}$.
The values are based on least-squares fits with equal weights, slightly
modified in two cases to keep the fit monotonic across the metallicity
range of our sample. For [N II], we adopt a flat value EW=1 \AA\ 
for (O/H)$<$0.5 solar. These forms are for interpolation purposes,
and their limitations appear from the fact that they do not all 
approach zero line strength at zero metallicity. Crudely speaking,
wind-dominated lines would have a quadratic dependence on 
metallicity until saturation sets in, since the mass-loss rate and
fraction of the mass in the right ionic state each depend on
metallicity, while mostly photospheric lines (such as P V)
are more nearly linear in strength with the abundances.

Strictly speaking, these relations apply only to galaxies with a
long (even if episodic or weak) history of star formation, since
the elements involved in these lines come from very different
stellar processes. For example, oxygen should be enriched
rather quickly, coming from massive stars, while carbon will
take longer coming from intermediate masses, with the dominant sources
nitrogen still somewhat ambiguous. At best, these relations could
be taken seriously for single elements in high-redshift galaxies,
and can in fact be used to test for the differential enrichment
history of such systems, in much the same way that the behavior of
Si IV and C IV lines with redshifts has been used to infer such differences
by Mehlert et al. (2002), who derive a relation between C IV EW
and metallicity consistent with our IUE analysis included in Fig. 6.  

\subsection {Far-UV spectra and the stellar mix}

Beyond the stellar-wind lines, the greatest difference among all these far-UV 
spectra is the flux excess
in NGC 604 from about 912-940 \AA , above what is seen in any of the
other objects either of higher or lower metallicity. Small-number statistics
in the massive stars might be expected to be more important for the
less rich systems, but in M33 it is these which have the same
spectral shape as the more luminous and distant systems, leaving NGC 604
as the odd one out. Previous work on its stellar
content indicates that there was a strong peak in its star-forming
history about 3 Myr ago, recent enough to affect the mix of stars contributing
in this spectral range.
We have examined additional UV data on the stellar
content of these M33 H II regions to see how NGC 604 might be different.

The stellar population in NGC 604 has drawn considerable attention, in being
one of the brightest star-forming complexes in the Local Group while
offering a dramatic structural contrast to the dominant, compact stellar
cluster in 30 Doradus. Bruhweiler, Miskey, \& Smith Neubig (2003) combined mid-UV
WFPC2 images with a wide-slit STIS spectrum to extract individual
spectra of the brightest members. They note that the ten most UV-luminous
stars will dominate the integrated spectral features from NGC 604, and that
two of these are located $\approx 0.3$ magnitude above the usual 
120-solar-mass limit. Specifically, the ``top ten" stars contribute 46\% of
the total measured flux just longward of Lyman $\alpha$.
Mid-UV STIS images (central wavelengths 1820-2700 \AA )
obtained in support of a slitless-spectroscopy program
by J. Mais-Apellaniz show, via the finer pixel sampling, that there are 
additional spatially resolved companions to each
of the brightest stars identified by Bruhweiler et al., although none so
bright as to bias the measured colors or magnitudes. NGC 604 also
contains significant numbers of WR stars, although with a WR/O
star ratio near 0.1 rather than the 0.3 seen in, for example, NGC 595
(Drissen, Moffat, \& Shara 1993).

To compare the populations in these H II regions as they affect the integrated
far-UV spectra, we produced color-magnitude arrays for each based on the 
archival WFPC2 data in F170W and F555W (as well as intermediate-wavelength
data when available). 
Lacking imagery in the far-UV band itself, we use the mid-UV properties
as proxies to at least identify the hottest luminous stars.
Reddening corrections do not enter for our
immediate purposes, since we need to know only how many stars contribute
to the UV flux and extinction does not
varying greatly between 1700 and 1100 \AA\ . Bruhweiler et al. find,
for stars in NGC 604, that the 1100-\AA\  extinction is about 1.5 times
that at 1700 \AA . 
Likewise, we are interested here only in stars selected from UV flux, so the 
diagrams
sample only those stars well detected in the mid-UV data.
These observational HR diagrams are shown in Fig. 7.
We include only stars without serious crowding issues, measured within
0.3" radii. The samples include stars to about $m_{170}=19$ on the STMAG
scale (based on flux per unit wavelength), although the completeness
varies considerably even at $m_{170}=18$ because of differences in
crowding; for NGC 604, the cumulative counts with flux suggest
statistical completeness only above $m_{170}=17.0$. The brightest stars in the 
rich population of NGC 604
are 1.5-2 magnitudes brighter than found in any of the other associations.
The color range among UV-bright stars in NGC 588 is smallest, extending
from the blue envelope near $m_{170} - m_{555} = -3.5$ redward only
to -2.0, while all the other regions have the diagram populated to
$m_{170} - m_{555} = 0$, and in NGC 604 to 2.4. This may in part be a reddening
issue, since the stellar distribution in NGC 588 is more compact,
with less scope for differential reddening across the association,
than the others. Indeed, dust lanes are prominent in the continuum
images of NGC 604 in both the optical continuum and UV/optical colors;
it is clear observationally that the UV reflects only the 
least-extinguished stars in this object. The observed color distribution
in NGC 604 is rich in the bluest stars, but these stars are also
represented with similar color in NGC 588 and 595. A more important
difference is that the richer population in NGC 604 includes several stars 
with F170W magnitude brighter than seen in any of the other clusters.
While richness effects mean that the brighter star-forming regions
will have brighter first-ranked stars, the studies referenced above
suggest that NGC 604 has undergone a distinct burst of star formation
about 3 Myr ago, younger than the other regions, an event which is
recent enough to leave its mark on the massive-star population. We
now focus on these brightest stars.

These data allow us to address how much the brightest stars dominate the mid-UV
flux in each case, incorporating the total F170W flux within the FUSE
aperture. Fluctuations in the bright population could affect the
integrated spectra of the lower-luminosity, sparser regions more strongly
simply from statistics, but the detailed recent history of star formation
will enter through aging as well. Given the temperature range of these
stars ($3-4 \times 10^4$ K following Bruhweiler et al.), the hottest stars
should be more dominant at the shorter wavelengths of the FUSE data.
It is not clear that all the far-UV flux is from direct starlight;
Hill et al. (1995a,b) and Malamuth et al. (1996) showed evidence 
that a significant fraction of mid-UV light
from similar systems is scattered. The HST F300W image of NGC 604 
shows reflection nebulosity, and smoothed versions of the
mid-UV STIS images match its morphology, indicating that scattering
important at shorter wavelengths as well, as would
be expected for a roughly $\lambda^{-4}$ Rayleigh behavior. 
Since the FUSE aperture is not much larger than the stellar
distribution in NGC 604, the widths of narrow absorption features
are not a sensitive test of whether scattering is important at these
wavelengths. We do see a role for scattering around some of the
brightest stars in the HST STIS image at 2400 \AA\ , from analysis of
the point-spread functions of stars. This is easier to interpret than
the distinct emission and absorption structures, since reflected
continuum can be confused with emission from the weak [O II] doublet
near 2471 \AA\ . Some of the stars match the nominal PSF closely,
while two of the brightest ones exhibit excess light from 0.3-1.0" from the
core. This excess contains as much of 28\% of the mid-UV light in the
brightest case. These data leave open the possibility that scattered light 
is important at the shorter far-UV wavelengths.

We therefore 
bracket the total
UV flux between the sum of detected stars and the large-aperture sum.
For larger fractions of scattered light, the brightest individual stars
are more dominant, since the total number of stars producing the
observed light is smaller. Cumulative distributions of F170W
magnitude are shown in Fig. 8, along with simple geometric estimates
of corrections for crowding (which are small, but underestimate the actual
effect where stars are more clumped than random within the H II region).
Accumulating the flux from the bright stars, we find that only
about 20\% of the total flux within the FUSE
aperture at 1700 \AA\  comes directly from the stars, less than half
of what Bruhweiler et al. (2003) find at 1200 \AA\  within a narrow
slit. This may be a sign that scattering is important, since much
of the scattered light we see at mid-UV wavelengths in the HST
imagery is on larger scales than this. In contrast, as a fraction of the
light {\it from detected stars}, the brightest ten (a good approximation to
the stars brighter than any found in the other H II regions) contribute 
about 40\%, more in line with the Bruhweiler et al. results. This
also makes sense for these stars being able to affect the overall
spectrum. However, the high temperature needed for the excess component
in NGC 604 means that we are seeing a difference in history rather
than simply small-number statistics in what stars appear at a
given time.

\subsection{Stellar populations}

In general, metallicity will be manifested in the composite spectra
both directly, through photospheric and wind lines, and indirectly,
as the evolutionary tracks of stars change with abundances. Effects
on the initial-mass function are too small to see at the abundances
found in the M33 disk, as shown by Malamuth et al. (1996) for some of the
same H II regions we observed. They suggest that the excitation trends
seen in the associated ionized gas result from the different
emergent ionizing fluxes for stars at various metallicities. However,
different evolutionary histories are still implied by the strong
changes in wind properties seen with metallicity. 

Comparison of the our spectra suggests that very recent events in
the star-formation history do have observable impact in the
far-UV, as exemplified by NGC 604. Its mix includes light from
a greater fraction of hot stars than the other systems,
which fits with the other properties of this region in suggesting
a burst so recent that even in the far-UV, it does not look 
like constant star-formation (in this case, ages $\approx 3 $ Myr).
In general, the timescales may be metallicity-dependent;
Robert et al. (2003) show that evolved O I/III and B I/III stars appear later
at lower abundances.

\section{Summary}

We have used FUSE spectra of star-forming regions in nearby galaxies,
whose gas-phase metallicities range from 0.05-0.8 solar, to explore the utility of
far-ultraviolet spectra in measuring the abundances in star-forming galaxies,
as well as to probe the massive-star populations in these galaxies. The
absorption lines from radiatively-driven winds prove to be very sensitive
to metal abundance; all six strong and unblended species (including C IV from
archival IUE data) have a strong, monotonic metallicity dependence. For 
N IV, C III, and P V, the relation between straighforward equivalent-width values
and oxygen abundance from emission-line spectra is closely linear, suggesting that
these lines will be useful in tracing the chemical history of galaxies from
$z=3-4$, beyond which the Lyman $\alpha$ forest makes even composite spectra
progressively less informative. 

The continuum of NGC 604 departs from the uniform shape of the other objects
below 950 \AA . After considering the effects of small-number statistics
among the massive stars in these objects, we conclude that this difference
probably traces to a discrete burst of star formation $\approx 3$ Myr ago in
NGC 604. This region had been considered by several previous studies to
have hoisted such a burst, on grounds of both morphology of the gas and
fitting of the H-R diagram.

\acknowledgments
B.-G. Andersson was helpful in understanding some of the issues
in scheduling and data analysis from FUSE.
We thank Alessandra Aloissi and her collaborators for providing their summed
FUSE spectrum of I Zw 18.
We also acknowledge the community service provided by
Steven McCandless in making his H$_2$OOLS compilation of data and
routines available. Dick Tipping patiently explained some of the intricacies
of the H$_2$ spectrum on several occasions.
This work was supported by NASA through FUSE GI grant NAG5-8959.
We also made use of the MAST archive system in retrieving data from
HST, IUE, and UIT. We thank the referee, Claus Leitherer, for a detailed,
expeditious, and helpful critique.

\clearpage
\figcaption
{Locations and stellar distributions in the regions observed
in M33. The background image is UIT image FUV-0496 (1500 \AA\ ), with
ellipses corresponding to galactocentric radii in the inner disk
of M33. Each cutout from HST mid-UV images is 45 arcseconds square
with north at the top, with an outline of the FUSE large science
aperture at the recorded position for each observation. These regions
range from single, compact associations (as in NGC 588) to the
extended and multiple collections in NGC 592 and 604. The dispersion
direction for these observations runs ENE-WSW; the limiting spectral
resolution is set by the distribution of the far-UV starlight in
this direction, which is a factor only for NGC 604. The images
are displayed with an offset logarithmic intensity scale. The
ellipses indicate distance from the nucleus in the disk plane,
taking the geometric parameters for this part of the disk (within
the inner non-warped region) from the optical fits by de Vaucouleurs (1959)
and the H I fits of Corbelli \& Schneider (1997). We adopt a distance of
850 kpc, in the middle of the range of distances from Cepheids (Lee et al.
2002), the tip of the red-giant branch (Kim et al. 2002), and
planetary nebulae (Kuzio et al. 1999).
\label{fig1}}

\figcaption
{FUSE aperture location and size for the NGC 5461 and 7714
observations, superimposed over the shortest-wavelength HST observation
available. The ACS image of NGC 5461, from program 9490 led by
K. Kuntz does not include the entire
FUSE aperture.
An inset shows the multiple luminous clusters in the starburst
nucleus of NGC 7714, observed by Windhorst et al. under program 9124.
As for the cutouts in Fig. 1, each image section
spans 45 arcseconds with north at the top.The dispersion direction for
NGC 5461 runs SSE-NNW, while for NGC 7714 it is ENE-WSW.
\label{fig2}}

\figcaption
{FUSE spectra of NGC 604 (heavy line) and I Zw 18, shown with 0.1-\AA\  pixels
in the heliocentric velocity frame, to reduce clutter from
the H$_2$ absorption features. Fluxes have been scaled up by
$10^{14}$ erg cm$^{-2}$ s$^{-1}$ \AA $^{-1}$ for NGC 604,
and by $2 \times 10^{14}$ for I Zw 18.
The numerous H$_2$ features (of which the most prominent
are marked by ticks below the spectra)
are from Milky Way gas. Features in NGC 604 are slightly blueshifted
(240 kms$^{-1}$, about 0.8 \AA  at 1000 \AA ), while the redshift of
I Zw 18 is $cz=751 $ km s$^{-1}$ for a typical wavelength shift of 
+2.4 \AA . Atomic interstellar absorption lines are marked at each 
redshift by vertical lines above the spectrum, where the shorter line
is for I Zw 18. Angled symbols near the top indicate the zero-redshift
locations of important
stellar-wind lines. The continuum of NGC 604 is higher below about
940 \AA . While the interstellar lines are nearly as strong in 
I Zw 18 as the much higher-metallicity disk of M33, the stellar wind
features are substantially weaker. The continuum level in NGC 604
is significantly higher shortward of about 955 \AA , and P Cygni emission
redward of the $\lambda 1037$ line is prominent in NGC 604 but
not in I Zw 18.
\label{fig3}}

\figcaption
{Comparison of the FUSE spectra of NGC 604 and I Zw 18, as in Fig. 3,
now with Galactic molecular absorption removed and both spectra
plotted in the emitted wavelength frame. Wind and interstellar lines are
marked as before; some unpatched foreground absorption remains.
Interstellar absorption features intrinsic to the surrounding 
galaxies stand out by matching in both spectra.
This comparison shows the difference in both wind absorption and
P Cygni between the abundances of I Zw 18 (O/H about 0.02 solar)
and NGC 604 (0.4 solar). Both effects are clear for O VI, C III, and N II.
\label{fig4}}

\figcaption
{Comparison of the H$_2$-corrected spectrum of NGC 604 to NGC 7714,
in the emitted frame as in Fig. 4. Stronger absorption is prominent
in the blue wings of O VI and C III, and in the overall profiles of
N II and Si III/IV. The NGC 7714 data have been boxcar-smoothed 
by 0.7 \AA\  and scaled by a factor $10^{14}$. 
\label{fig5}}

\figcaption
{Equivalent widths of stellar-wind lines in the spectra of star-forming regions.
Each shows a strong metalliity dependence, here quantified using the
traditional emission-line results for O/H. Typical errors for the far-UV
lines are $\pm 0.2$ \AA\ , with the IUE spectra used for C IV accurate to about
$\pm 1$ \AA . The three fainter M33 regions are averaged into single
points for each transition. The quadratic interpolation functions
with coefficients listed in Table 3 are overplotted as guides, where we
take a flat value EW=1 \AA\ for N II below (O/H)=0.5. 
\label{fig6}}

\figcaption
{Observed color-magnitude arrays for H II regions in M33,
derived from archival WFPC2 images. The two brightest objects in
NGC 604 have less certain colors due to saturation in the
F555W images. These are shown in the STMAG system, in which zero
color index corresponds to constant F$_\lambda$. Since we are 
interested in which stars contribute to the far-UV flux, no reddening 
corrections have been applied. The stars of interest are so blue that
red-leak corrections in the F170W filter are negligible for our purposes.
\label{fig7}}

\figcaption
{Cumulative star counts as observed at 1700 \AA\  for the M33 H II
regions, including stars within the FUSE apertures. In each region,
the upper curve includes a simple corection for crowding, made by assuming
that the fainter stars are uniformly distributed through the populated
region in each association.
\label{fig8}}

\clearpage

\begin{table*}
\begin{center}
\begin{tabular}{lcccrcl}
\tableline
\tableline
Object   &   RA     &    Dec  &  $cz$ & Exposure & O/H &  source\cr
   & $\alpha_{2000}$ & $\delta_{2000}$ & km s$^{-1}$ & seconds & (solar units) & \cr
NGC 588  & 01 32 45.50 & +30 38 55 & -174  &  5220  & 0.41 & Vilchez et al. 1988 \cr
NGC 592  & 01 33 12.27 & +30 38 49 & -162  &  3965  & 0.48 & Interpolated \cr
NGC 595  & 01 33 33.60 & +30 41 32 & -178  &  7123  & 0.56 & Vilchez et al. 1988 \cr
NGC 604  & 01 34 32.50 & +30 47 04 & -226  &  7151  & 0.66 & Vilchez et al. 1988 \cr
NGC 5461 & 14 03 41.30 & +54 19 05 &  298  &  5189  & 0.68 & Luridiana et al. 2002 \cr
NGC 7714 & 23 36 14.0  & +02 09 19 & 2798  &  6023  & 0.81 & Gonzalez-Delgado et al. 1995 \cr
I Zw 18  & 09 34 02.30 & +55 14 25 &  751  & 95097  & 0.05 & Izotov et al. 1999 \cr
\tableline
\tablecomments{O/H is  in solar units, converted when necessary assuming a solar
value of 12+log O/H=8.60 following Allende Prieto, Lambert, \& Asplund (2001)}
\end{tabular}
\tablenum{1}
\caption{
FUSE Targets and Properties \label{tbl1}}
\end{center}
\end{table*}

\clearpage

\begin{table*}
\begin{center}
\begin{tabular}{lcccccc}
\tableline
\tableline
Object & N IV 955  &  C III 977 &   N III 991 &    N II 1083  &   P V 1122 &
 C IV 1549\cr
I Zw 18  & 0.27: &  0.47 &  $<0.2$ ($<0.2$) & 1.11 (1.02) &  0.27 & 2.0:\cr
M33 avg  & 1.26  &  1.19 &   ... (...)      & 0.86 (0.24) & 0.66  & 8.4 (7.8) \cr
NGC 604  & 2.17  &  1.83 &   1.77 (0.59)    & 1.50 (1.33) & 0.89  & 9.49 (7.97) \cr
NGC 5461 & 2.11  &  1.56 &   2.40 (1.87)    & 2.19 (1.92) & 0.90  & 7.82 (6.03)\cr
NGC 7714 & 3.26  &  2.56 &   3.68 (2.36)    & 2.52 (2.52) & 1.01  & 10.1 (10.2)\cr
\tableline
\tablecomments{All values are in \AA\  in the emitted frame. Parenthesized
values include the emission component of a P Cygni profile.}
\end{tabular}
\tablenum{2}
\caption{
Equivalent Widths of Stellar Wind Lines \label{tbl2}}
\end{center}
\end{table*}

\clearpage

\begin{table*}
\begin{center}
\begin{tabular}{lccccccc}
\tableline
\tableline

Transition & $a_1$ & $a_2$ & $a_3$ &  $c_1$ & $c_2$ & $c_3$ & $c_4$ \\
N IV &  0.30 & -0.656 &   5.2201 &  0.063  & 0.0958 &  -5.8314 & 20.8804 \\
N III & 0.05 & -1.079 &   6.5539 &  0.082  & 0.0763 &  -0.0794 & 26.2157 \\
N II &  1.28 & -4.029 &   7.0447 &  0.286  & 0.0710  & -19.7753 & 28.1790 \\
C IV &  1.14 & 19.210 & -10.8494 &  0.885  & -0.0461 &  418.6269 & -43.3975 \\
C III & 0.50 & -0.502 &   3.6120 &  0.070  & 0.1384  &  -6.9750 & 14.4481 \\
P V &   0.22 &  0.910 &   0.1037 & -4.386  & 4.8224  &  0.7356 & 0.41473 \\
\tableline
\tablecomments{Entries are coefficients of forward and inverse quadratic
fits as listed in the text, when equivalent widths are in \AA\  and
O/H is in solar units.}
\end{tabular}
\tablenum{3}
\caption{
Quadratic Fits for Line Strength versus Metallicity \label{tbl3}}
\end{center}
\end{table*}

\end{document}